\documentclass[12pt]{article}
\usepackage{graphicx}
\pdfoutput=1

\def\Title#1{\begin{center} {\Large #1 } \end{center}}
\def\Author#1{\begin{center}{ \sc #1} \end{center}}
\def\Address#1{\begin{center}{ \it #1} \end{center}}

\newenvironment{Abstract}{\begin{center}{\bf Abstract}\end{center} \bigskip \begin{quotation}  }{\end{quotation}}
\newenvironment{Presented}{\begin{quotation} \begin{center} 
             PRESENTED AT\end{center}\bigskip 
      \begin{center}\begin{large}}{\end{large}\end{center} \end{quotation}}

\def\beq{\begin{equation}}
\def\eeq#1{\label{#1}\end{equation}}
\def\eeqn{\end{equation}}

\def\beqa{\begin{eqnarray}}
\def\eeqa#1{\label{#1}\end{eqnarray}}
\def\eeqan{\end{eqnarray}}

\let\bar=\overbar

\def\Dslash{\not{\hbox{\kern-4pt $D$}}}
\def\dslash{\not{\hbox{\kern-2pt $\del$}}}

\def\msb{{\bar{\ssstyle M \kern -1pt S}}}

\begin{document}
\begin{titlepage}

\vfill


\Title{Results on top-quark physics from the CMS experiment}
\vfill
\Author{S. Tosi\\ on behalf of the CMS Collaboration}  
\Address{Institut de Physique Nucl\'eaire de Lyon, Villeurbanne,
  69100, France}
\vfill


\begin{Abstract}
The most recent results on top-quark physics reported by the CMS
experiment at the Large Hadron Collider (LHC) are presented in this talk.
The results are based on a data sample of about 36 pb$^{-1}$ of data collected
during 2010 at a $pp$ center-of-mass energy of 7 TeV.
\end{Abstract}

\vfill

\begin{Presented}
The Ninth International Conference on\\
Flavor Physics and CP Violation\\
(FPCP 2011)\\
Maale Hachamisha, Israel,  May 23--27, 2011
\end{Presented}
\vfill

\end{titlepage}
\def\thefootnote{\fnsymbol{footnote}}
\setcounter{footnote}{0}
%


\section{Introduction}

The top quark has a special place in the Standard Model (SM). It is the
heaviest elementary particle known to date, its mass being very close
to the scale of electroweak symmetry breaking. As such it plays a
special role in many models of new physics beyond the SM.

In the SM, the top quark decays almost exclusively to a $Wb$ pair.
At the LHC~\cite{LHC}, the top quark is predominantly produced in pairs, via the
strong interaction, with a total cross-section of around 158 pb, at the NLO. From an
experimental point of view, $t\bar{t}$ pairs can be classified in
three channels according to the decay of the two $W$ bosons originated by the top and
anti-top decays: dileptonic channel, when both $W$ decay leptonically;
all-hadronic channel, when both $W$ decay to quarks; semi-leptonic channel, when one
$W$ decays to leptons and the other to quarks. The three channels have a
 branching fraction of about 10, 48 and 42 $\%$, respectively.

An additional production mechanism is single-top production, via the 
electro-weak interaction. Three possible channels exist with a total
cross section of around 78 pb.

The top quark has a paramount importance at the LHC. Final states
originated from top decays typically include jets, leptons and missing energy, hence typically involve
almost all subdetectors and allow thorough tests of the performances of the
detector. In addition, studies of top-quark production and properties represent an
important tool to verify SM predictions and QCD calculations in the LHC environment. 

Several extensions of the SM foresee a preferential
coupling to the third generation and in particular to the top-quark
sector. For example, new resonances may exist decaying to
top-antitop pairs.

\section{The LHC and CMS}

The LHC started to deliver $pp$ collisions at the end of 2009. Following
two periods of data taking at a $pp$ center-of-mass energy of 900 GeV and 2360 GeV, during year 2010
collisions at 7 TeV have been recorded. The total integrated
luminosity in 2010, on which the results shown here are based, was about 36 pb$^{-1}$. During 2011 the LHC has already delivered
more than 1 fb$^{-1}$ of data at a center-of-mass energy of 7 TeV.

The collisions produced by the LHC are recorded with high efficiency
by the CMS detector, which is descibed in detail elsewhere~\cite{CMS_detector}.

\section{Ingredients for top physics}

The reconstruction of final states of top decays typically involves
several basic objects at the same time: leptons, jets, missing energy.

At the trigger level, events containing top quarks are selected by
requiring the presence of at least one
electron or one muon. The minimum $E_T$ of the electrons was requested
to exceed 10 GeV in the beginning of the data taking period, this threshold being raised with the increase of
the event rate to 22 GeV. The efficiency of the electron trigger requirements
is above $98\%$. Similarly, the minimum $p_T$ of muons was requested
to exceed 9 GeV/c, the threshold being raised up to 15 GeV/c at higher
event rates. The efficiency of the muon trigger requirements is above
$92\%$.

Jets and missing energy are reconstructed using the ``particle flow'' algorithm, where
the information from all sub-detectors is combined to determine the
particle content of the events. This greatly improves on the
performances of jet identification and resolution with respect to a
reconstruction only based on energy deposits in the calorimeters. The
energy scale uncertainty for jets typically varies between 3 to $5\%$
depending on the $E_T$, and the uncertainty on the
resolution is around $10\%$.

In events containing top pairs, two jets are originated from the
hadronization of a $b$ quark. Several algorithms have been developed to
identify the $b$-flavor of jets, which exploit various properties
of $b$-hadrons, for instance the presence of a displaced secondary
vertex or of secondary non-isolated leptons. These algorithms cover a broad range of signal
efficiency and mis-identification rates and are chosen by each data analysis
according to the specific needs. 

\section{Inclusive cross-section measurement}

\begin{figure}[htb]
\centering
\includegraphics[width=0.6\textwidth]{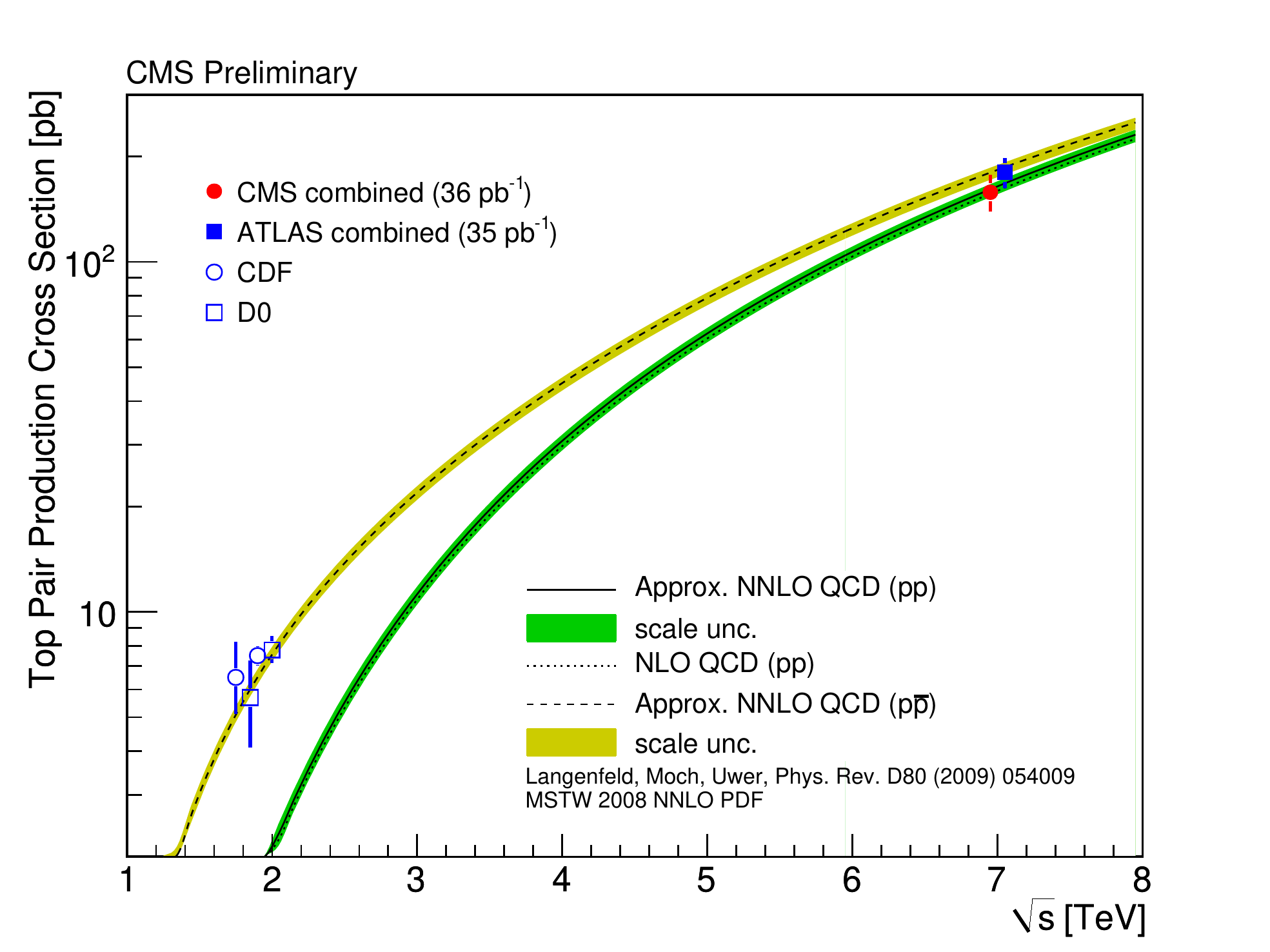}
\caption{Top pair production cross section as a function of $\sqrt{s}$, for both $p\bar{p}$ and $pp$ collisions. Data points are slightly displaced horizontally for better visibility. Theory predictions at approximate NNLO are obtained using HATHOR~\cite{Hathor}. The error band of the prediction corresponds to the scale uncertainty.}
\label{fig:xsec_combo}
\end{figure}

The inclusive cross section of $t\bar{t}$-pair production has been
measured by CMS using both the semi-leptonic and the dileptonic channels. In both
cases, electrons and muons have been considered. The cross section
with the semileptonic channel has been measured both with and without
the usage of $b$-tagging algorithms. In the latter case~\cite{Chatrchyan:2011ew}, the signal
content is extraced by means of a binned likelihood fit to the missing energy
and the invariant mass of the three jets yielding the highest summed
$p_T$. In the former case~\cite{TOP-10-003}, a binned likelihood fit to the invariant
mass of all objects pertaining to the secondary vertex is performed,
where systematic uncertainties are included as nuisance parameters in
the fit. The cross section was mesured to be $(173\pm
14(stat)^{+36}_{-29}(syst)\pm7(lumi))$ and $(150\pm9(stat)\pm17(syst)\pm6(lumi))$ pb, respectively.
In the dileptonic channel, the signal content is extracted by counting
the events in jet multiplicity bins and the cross section is measured
to be $(168\pm18(stat)\pm14(syst)\pm7(lumi))$ pb~\cite{Chatrchyan:2011nb}. Combining the semi-leptonic and the
dileptonic cross-section measurements, properly taking into account
the correletions, CMS obtains $(158\pm19)$ pb~\cite{TOP-11-001}, consistent with
the SM expectation (Fig.~\ref{fig:xsec_combo}).

\section{Single top}

The first measurement of the $t$-channel single top production cross
section in pp collisions at $\sqrt{s}$= 7 TeV has been performed by
CMS using two analysis methods~\cite{Chatrchyan:2011vp}. The first method makes use of a template fit
to two angular variables sensitive to the $t$-channel single top
production, the pseudorapidity distribution of the light jet
accompanying the top quark and the
angle between this jet and the lepton issued in the top decay chain.
The second method makes use of a multivariate boosted decision tree technique, combining 37
event-shape and kinematic variables. An evidence exceeding three
standard deviations has been obtained with both methods, and the
 cross section is measured to be $(83.6\pm29.8(stat+syst)\pm3.3(lumi))$ pb.

\section{Measurement of the top-quark mass}

The mass of the top quark is a fundamental parameter in the SM and it
affects predictions of SM observables via radiative corrections.
Several methods have been developed to measure the top-quark mass. CMS
used improved versions of the matrix weighting technique~\cite{top_mass_d0} and
the fully kinematic method~\cite{top_mass_cdf}. Combining the results yielded by the
two methods, the mass was measured to be $(175.5\pm4.6(stat)\pm4.6(syst))$ GeV/c$^2$~\cite{Chatrchyan:2011nb}. 

\section{Search for resonances decaying to  $t\bar{t}$ and measurement of the
  charge asymmetry}

New Physics can manifest itself in several ways in top-pair
production~\cite{NPTop}, often with intermediate new resonances,
generically referred to as $Z'$,
that decay to  $t\bar{t}$ pairs. 

\begin{figure}[htb]
\centering
\includegraphics[width=0.6\textwidth]{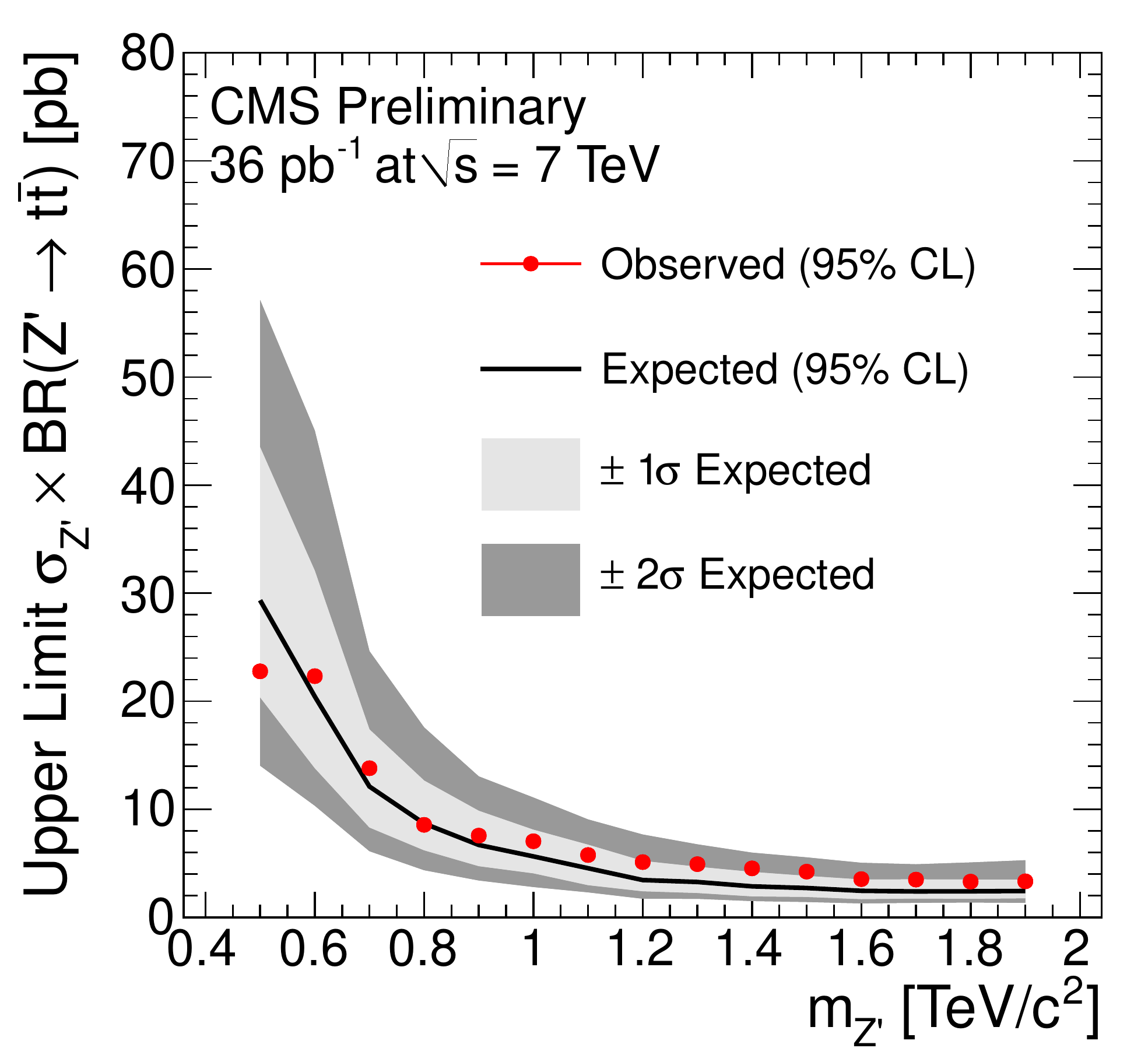}
\caption{Expected and observed 95 $\%$ C.L. upper limits for $\sigma(pp \to Z') \times BR(Z' \to t
  \bar{t} )$ for 36 pb$^{-1}$ of data as a function of the $Z'$ mass.}
\label{fig:ttmass}
\end{figure}

A direct search for narrow $Z'$ resonances decaying to $t\bar{t}$ pairs has
been performed using the semi-leptonic channel~\cite{TOP-010-007}. A template likelihood
fit to the  $t\bar{t}$ invariant mass has been used. No signal has been
observed and upper limits on the $Z'$ production cross section have been
derived as a function of the $Z'$ mass, as shown in Fig.~\ref{fig:ttmass}.

\begin{figure}[htb]
\centering
\includegraphics[width=0.6\textwidth]{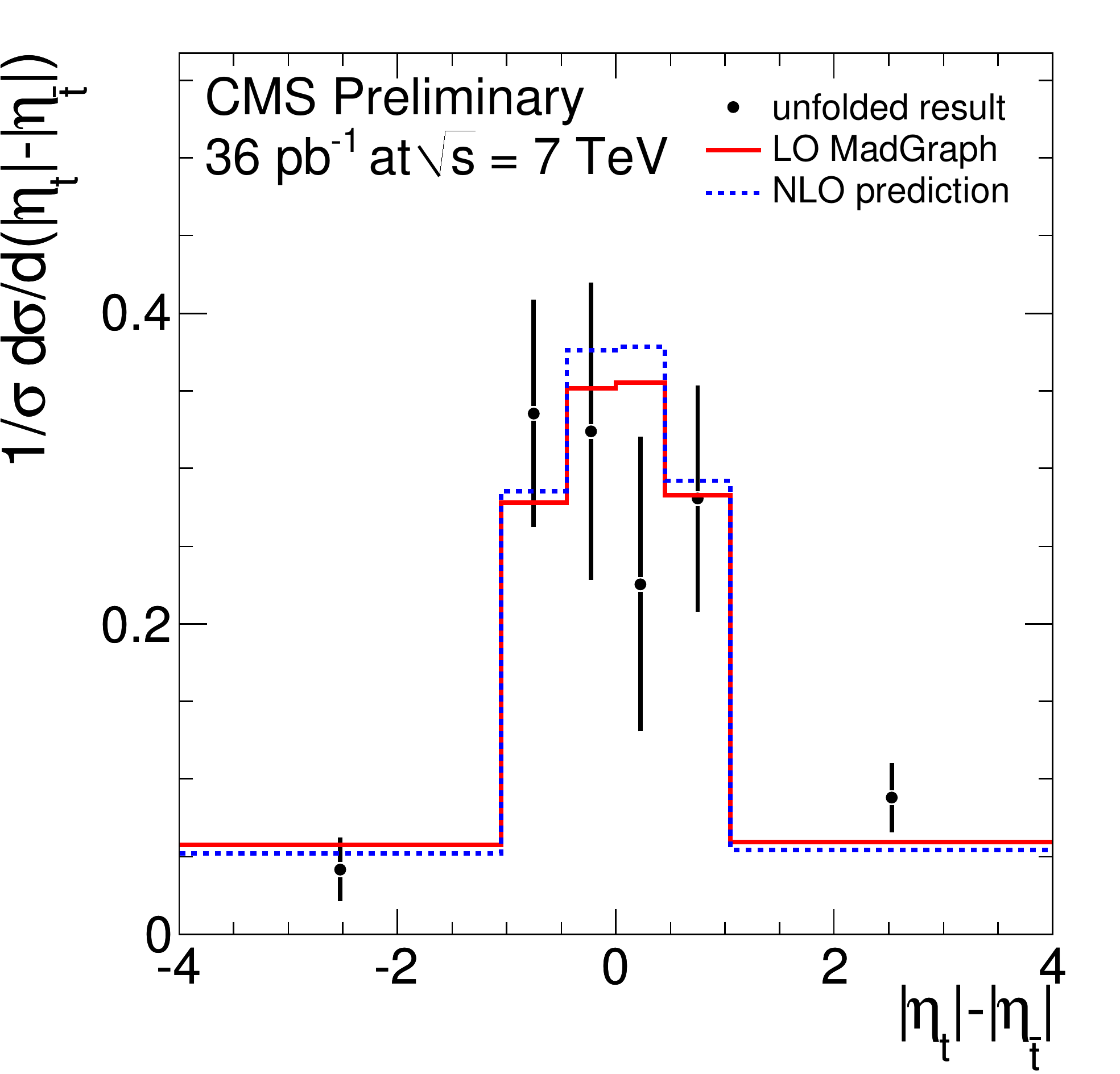}
\caption{ Distribution of the unfolded  $|\eta_t|-|\eta_{\bar{t}}|$
 spectrum. The shown theory curves are the prediction from the MADGRAPH generator and a NLO computation~\cite{asym_theory}.}
\label{fig:asymmetry}
\end{figure}

Broad resonances can escape identification in a measurement of the invariant
mass spectrum. However their presence can be inferred in other ways.
One possibility is the measurement of the charge asymmetry. In the SM, top
pairs are produced in a symmetric state, as far as the color charge is concerned,
at the LO. At the NLO, a small asymmetry is present, which can be
enhanced if the $t\bar{t}$ pair is originated by the decay of a heavy
resonance. At the Tevatron experiments, where the initial state is
$p\bar{p}$, a charge asymmetry yields a forward-backward
asymmetry: the Tevatron experiments indeed reported a deviation of the
forward-backward asymmetry from SM expectations by around 2 standard
deviations~\cite{Tevatron_chargeasym}. At the LHC, where the initial state is $pp$, a charge
asymmetry can induce a difference of absolute pseudo rapidities of top
 and anti-top quarks, $|\eta_t|-|\eta_{\bar{t}}|$ (Fig.~\ref{fig:asymmetry}). CMS measures an asymmetry of $(0.060\pm0.134(stat)\pm0.026(syst))$, consistent
with the SM expectations~\cite{TOP-010-010}.

\section{Conclusion}

The CMS experiment has already reported several interesting results on
top-quark physics. So far, all results are consistent with the SM
predictions.  A rich program of measurements of 
differential distributions and properties of the top quark is in place, including 
the usage of additional channels. A lot more data have already been
recorded and exciting 
results are expected soon.



\begin{thebibliography}{99}

\bibitem{LHC}
L. Evans and P. Bryant (editors), ``LHC Machine'', JINST 3 (2008) S08001. 

\bibitem{CMS_detector}
CMS Collaboration, ``The CMS experiment at the CERN LHC'', JINST 0803 (2008) 
S08004. 

\bibitem{Chatrchyan:2011ew}
  S.~Chatrchyan {\it et al.}  [CMS Collaboration],
  ``Measurement of the Top-antitop Production Cross Section in pp Collisions at
  sqrt(s)=7 TeV using the Kinematic Properties of Events with Leptons and
  Jets,''
  arXiv:1106.0902 [hep-ex].

\bibitem{TOP-10-003}
CMS Collaboration, ``Measurement of the $t\bar{t}t$ Pair Production Cross Section at $\sqrt{s}$ = 7 TeV 
using b-quark Jet Identification Techniques in Lepton + Jet Events'', CMS TOP-11-001 
(2011). https://twiki.cern.ch/twiki/bin/view/CMSPublic/PhysicsResultsTOP11001. 


\bibitem{TOP-11-001}
CMS Collaboration, ``Combination of top pair production cross sections in pp collisions at 7 TeV and comparisons with theory'', CMS TOP-11-001 
(2011). http://cdsweb.cern.ch/record/1336491?ln=en

\bibitem{Hathor}
M. Aliev, H. Lacker, U. Langenfeld {\it et al.}, ``HATHOR: HAdronic Top and Heavy quarks 
crOss section calculatoR'', Comput.Phys.Commun. 182 (2011) 1034-1046, 
arXiv:1007.1327.

\bibitem{Chatrchyan:2011vp}
  S.~Chatrchyan {\it et al.}  [CMS Collaboration],
  ``Measurement of the t-channel single top quark production cross section in
  pp collisions at sqrt(s) = 7 TeV,''
  arXiv:1106.3052 [hep-ex].

\bibitem{top_mass_d0}
D0 Collaboration, ``Measurement of the top quark mass using dilepton events'', Phys. 
Rev. Lett. 80 (1998) 2063, arXiv:hep-ex/9706014. 


\bibitem{top_mass_cdf}
CDF Collaboration, ``Measurement of the top quark mass using template methods on 
dilepton events in proton antiproton collisions at $\sqrt{s}$ =1.96 TeV'', Phys. Rev. D73 (2006) 
112006, arXiv:hep-ex/0602008. 


\bibitem{Chatrchyan:2011nb}
  S.~Chatrchyan {\it et al.}  [CMS Collaboration],
  ``Measurement of the t t-bar production cross section and the top quark mass in the dilepton channel in pp collisions at sqrt(s) =7 TeV,''
  arXiv:1105.5661 [hep-ex].

\bibitem{NPTop}
See for example: S. Dimopoulos and H. Georgi Nucl. Phys. B193 150; S. Weinberg Phys. Rev. D13 974; L. Susskind Phys. Rev. D20 2619; C. T. Hill and J. Parke Phys. Rev. D49 4454; R. S. Chivukula et al. Phys. Rev. D59 075003; N. Arkani-Hamed, A. G. Cohen, and H. Georgi Phys. Lett. B513 232; N. Arkani-Hamed, S. Dimopoulos, and G. R. Dvali Phys. Lett. B429 263; L. Randall and R. Sundrum Phys. Rev. Lett. 83 3370. 

\bibitem{TOP-010-007}
CMS Collaboration, ``Search for Resonances in Semi-leptonic
Top-pair Decays Close to Production Threshold'', CMS-PAS-TOP-10-007. http://cdsweb.cern.ch/record/1335720?ln=en

\bibitem{Tevatron_chargeasym}
CDF Collaboration, ``Evidence for a Mass Dependent Forward-Backward Asymmetry in 
Top Quark Pair Production'', arXiv:1101.0034; D0 Collaboration, ``Measurement of the forward-backward production asymmetry of $t$ 
and $\bar{t}$ quarks in $pp \to t \bar{t}$ events'', DO note 6062-CONF (2010); CDF Collaboration, ``Forward-Backward Asymmetry in Top Quark Production in $pp$ Collisions at $\sqrt{s}$ = 1.96 TeV'', Phys. Rev. Lett. 101 (2008) 202001, arXiv:0806.2472. 

\bibitem{TOP-010-010}
CMS Collaboration, ``Measurement of the charge asymmetry in top quark
pair production with the CMS experiment'', CMS-PAS-TOP-10-010. http://cdsweb.cern.ch/record/1335714?ln=en

\bibitem{asym_theory}
P. Ferrario and G. Rodrigo, ``Massive color-octet bosons and the charge asymmetries of 
top quarks at hadron colliders'', Phys. Rev. D78 (2008) 094018, arXiv:0809.3354.  P. Ferrario and G. Rodrigo, ``Heavy colored resonances in top-antitop + jet at the LHC'', 
JHEP 02 (2010) 051, arXiv:0912.0687. 


\end{thebibliography}
\end{document}